# Comet C/2017 S3 (PanSTARRS): Outbursts and Disintegration

Short Title: Outbursts and Disintegration of Comet C/2017 S3 (PanSTARRS)

M.R. Combi[1], T. Mäkinen[2], J.-L. Bertaux[3], E. Quémerais[3], S. Ferron[4]
and R. Coronel[1]

[1]Dept. of Climate and Space Sciences and Engineering
University of Michigan
2455 Hayward Street
Ann Arbor, MI 48109-2143
*Corresponding author: mcombi@umich.edu

[2]Finnish Meteorological Institute, Box 503
SF-00101 Helsinki, FINLAND

[3]LATMOS/IPSL
Université de Versailles Saint-Quentin
11, Boulevard d'Alembert, 78280, Guyancourt, FRANCE

[4]ACRI-st, Sophia-Antipolis, FRANCE

Pages 18
Tables 3
Figures 3

**ABSTRACT**

The Solar Wind ANisotropies (SWAN) all-sky hydrogen Lyman-alpha camera on the SOlar and Heliospheric Observer (SOHO) satellite observed the hydrogen coma of comet C/2017 S3 (PanSTARRS) for the last month of its activity from 2018 July 4 to August 4 and what appears to have been its final disintegration just 11 days before its perihelion on August 15. The hydrogen coma indicated water production had a small outburst on July 8 at a heliocentric distance of 1.1AU and then a much larger one on July 20 at 0.8 AU. Over the following two weeks the water production dropped by more than a factor of ten after which it was no longer detectable. The behavior is reminiscent of comet C/1999 S4 (LINEAR) in 2000, which had a few small outbursts on its inbound orbit and a major outburst at a heliocentric distance of about 0.8 AU, which was close to its perihelion, followed by its complete disintegration that was documented by several sets of observations including SWAN. C/2017 S3 (PanSTARRS) however had a much larger water production rate than C/ 1999 S4 (LINEAR). Here we estimate the size of the nucleus of C/2017 S3 just before its final outburst and apparent disintegration was estimated using the total amount of water produced during its last weeks for a range of values of the refractory/ice ratio in the nucleus. We also determine the size distribution of the disintegrating particles as the comet faded.

Key Words: Comets; Cometary Atmospheres; Comet C/2017 S3 (PanSTARRS)

**Introduction**

The discovery of comet C/2017 S3 (PanSTARRS) was reported by Wainscoat et al. (2017) using the 1.8-m PanSTARRS telescope on September 23, 2017. Its orbit showed it would have a perihelion distance of 0.208 au on August 15, 2018. Visual magnitudes (Lehky et al. 2018; see also Yoshida 2018) show likely outbursts around July 4 and July 20, 2018, followed by a large drop in brightness of 2 to 3 magnitudes after that despite the comet's continued decreasing heliocentric distance on its way to perihelion. This would indicate a likely complete disintegration of the nucleus. Later reported visual magnitudes are likely just the dispersed dust cloud. According to the IAU Minor Planet Center (https://www.minorplanetcenter.net/) the comet's original semi-major axis, $a_0$, is negative or slightly hyperbolic, but within uncertainties yields a lower limit of 92500 AU or a value of $1/a_0$ of 0.0000108 putting it well into the normal dynamically new category (A'Hearn et al. 1995) coming from the Oort cloud.

Previous comets have had major outbursts, having thrown off fragments and a few have totally disintegrated. Comet C/1996 B2 (Hyakutake) had a major outburst event releasing both large fragments seen drifting anti-sunward over the following days (Harris et al. 1997; Rodionov et al. 1998; Desvoivres et al. 2000). The collection of smaller fragments released produced a halo of small fragments that greatly, but temporarily, increased the gas production rate, which returned to normal levels after a few days (Schleicher et al. 1998; Combi et al. 2005). Recently comet C2012 S1 (ISON) was widely observed because of its predicted perihelion distance of only 0.0124 AU, or ~2.7 solar radii from the center of the Sun. The nucleus did not survive its close passage to the Sun.

Comet C/1999 S4 (LINEAR) had been observed by a wide range of observatories and rather continuously over its whole apparition because of its predicted perihelion distance (Weaver et al. 2001; Bockelée-Morvan et al. 2000; Farnham et al. 2000; Mäkinen et al. 2001). Like 2017 S3, it was a dynamically new comet and also similarly completely disintegrated on the inbound leg of the orbit when it reached a heliocentric distance of ~0.8 AU.

**Observations and basic analysis**

The Solar Wind ANisotropies (SWAN) instrument on board the SOlar and Heliospheric Observatory (SOHO) satellite is a far ultraviolet all-sky camera sensitive to the hydrogen Lyman-α emission. It was designed for SOHO to be sensitive to the fluorescence excitation of the Lyman-$\alpha$ emission of the atomic hydrogen atoms that stream through the solar system from interplanetary space (Bertaux et al. 1995). Maps of the whole sky, nearly 4$\pi$ steradians, provide a 3-D image of the solar flux that leaves its signature in the loss of the streaming interstellar hydrogen that make up the interplanetary background. Because it is sensitive to H Ly$\alpha$ emission SWAN also serves as an excellent observatory for the fluorescence emission of the large hydrogen comae of comets that is produced by the photodissociation of $H_2O$, the typically most abundance volatile constituent of comets (Bertaux et al. 1998). As such, SWAN has observed over 60 comets in the past 21 years from which water production rates have been calculated (Combi et al. 2019). Because SOHO is located at the Earth-Sun L1 Lagrange point, comets of sufficient brightness in the entire sky can be observed in either the northern or southern hemisphere with none of the typical Earth horizon limitations from ground-based or low Earth orbit based observations. SWAN has only moderate exclusion zones around the location of the Sun as well as regions obscured by the spacecraft itself in the general direction of the Earth as seen from L1.

SWAN is now operated in a largely automatic mode scanning the entire sky with its 25 x 25 1 arc sec instrument field-of-view pixels every day. SWAN is in two parts, with one covering essentially the north heliographic hemisphere and the other covering the south. Images of comets are identified using their orbital elements. Water production rates are determined using our time-resolved-model (TRM), which is described in detail in the paper by Mäkinen and Combi (2005), using a combination of methods from Festou's (1981) vectorial model, the syndyname model of Keller and Meier (1976), and the Monte Carlo particle trajectory model of Combi and Smyth (1988a, 1988b). The H Ly$\alpha$ coma is typically captured in an 8-degree circular field of view where field stars are manually masked and the model fits both the comet's hydrogen distribution and the underlying interplanetary hydrogen background. Depending on a number of circumstances, including the concentration of background stars, the solar elongation angle, the brightness of the interplanetary H Lyα background, which is not a constant, and apparent cometary dust to gas ratio, comets with visual magnitudes brighter than magnitude 10-12 are usually detectable in SWAN so that water production rates can be calculated.

Water production rates are calculated for each image, but because of the filling time of the field of view by hydrogen atoms the production rates can represent an average over the previous 2-3 days, depending on the geocentric distance. However, if a comet is bright enough and spatially extended, a feature of the TRM, which simultaneously analyzes the various locations in the hydrogen coma in all images at the same time, is used. It deconvolves the temporal/spatial information inherent in the coma accounting for the time to produce H atoms by the photodissociation chain of $H_2O$ and OH as well as the transit time of H atoms in the coma. From this daily-average water production rates from the vicinity of the nucleus are calculated. See Combi et al. (2005) and Combi et al. (2014) for examples of its use.

SWAN observations of comet C/2017 S3 (PanSTARRS) were obtained from 4 July through 5 August 2018. The observational circumstances, g-factors, single-image water production rates and formal 1-σ uncertainties resulting from noise in the data and fitting procedure, are given in Table 1. Expected Total uncertainties resulting from a combination of calibration and model description and parameters are expected to be on the order of ~30%. The g-factors are calculated from the composite solar Lyα flux data taken from the LASP web site at the University of Colorado (http://lasp.colorado.edu/lisird/lya/). The value from the face of the Sun seen by the comet is taken from the nearest time accounting for the number of days of solar rotation between the Earth and comet locations. The shape of the solar Lyα line profile is taken from observation of Lemaire et al. (1998).

Table 1. SOHO/SWAN Observations of Comet 2017 S3 (PanSTARRS) and Water Production Rates

| Date (2018 UT) | r (AU) | Δ (AU) | g (s⁻¹) | Q (10²⁸ s⁻¹) | δQ (10²⁸ s⁻¹) |
|---|---|---|---|---|---|
| Jul 3.89 | 1.179 | 1.476 | 0.002542 | 1.399 | 0.07 |
| Jul 5.89 | 1.139 | 1.425 | 0.002544 | 1.543 | 0.03 |
| Jul 7.89 | 1.097 | 1.374 | 0.002574 | 1.974 | 0.03 |
| Jul 8.89 | 1.076 | 1.348 | 0.002575 | 1.476 | 0.04 |
| Jul 9.89 | 1.055 | 1.323 | 0.002603 | 1.679 | 0.15 |
| Jul 10.89 | 1.033 | 1.297 | 0.002603 | 1.440 | 0.04 |
| Jul 11.89 | 1.012 | 1.271 | 0.002604 | 1.338 | 0.07 |
| Jul 12.89 | 0.990 | 1.245 | 0.002605 | 1.428 | 0.13 |
| Jul 15.92 | 0.924 | 1.167 | 0.002635 | 1.539 | 0.26 |
| Jul 16.92 | 0.902 | 1.141 | 0.002636 | 1.938 | 0.17 |
| Jul 17.92 | 0.879 | 1.115 | 0.002663 | 3.633 | 0.36 |
| Jul 18.92 | 0.856 | 1.090 | 0.002664 | 4.295 | 0.89 |
| Jul 19.92 | 0.833 | 1.064 | 0.002665 | 5.611 | 0.09 |
| Jul 20.92 | 0.810 | 1.039 | 0.002690 | 6.197 | 0.12 |
| Jul 21.94 | 0.786 | 1.013 | 0.002691 | 5.778 | 0.11 |
| Jul 22.95 | 0.762 | 0.989 | 0.002717 | 4.662 | 0.08 |
| Jul 23.94 | 0.739 | 0.964 | 0.002718 | 3.598 | 0.17 |
| Jul 24.95 | 0.714 | 0.940 | 0.002719 | 3.093 | 0.07 |
| Jul 25.95 | 0.690 | 0.917 | 0.002743 | 2.118 | 0.08 |
| Jul 26.97 | 0.665 | 0.894 | 0.002744 | 2.150 | 0.25 |
| Jul 27.98 | 0.640 | 0.872 | 0.002765 | 1.509 | 0.09 |
| Jul 28.98 | 0.615 | 0.851 | 0.002766 | 1.174 | 0.13 |
| Jul 29.98 | 0.589 | 0.832 | 0.002788 | 1.231 | 0.29 |
| Aug 0.01 | 0.563 | 0.813 | 0.002789 | 1.060 | 0.23 |
| Aug 1.01 | 0.537 | 0.796 | 0.002790 | 0.590 | 0.42 |
| Aug 2.03 | 0.510 | 0.781 | 0.002804 | 0.411 | 0.48 |
| Aug 3.04 | 0.484 | 0.768 | 0.002805 | 0.376 | 0.32 |
| Aug 4.04 | 0.458 | 0.757 | 0.002806 | 0.594 | 0.62 |

Notes to Table 1. Date (UT) in 2018
r : Heliocentric distance (AU)
Δ: Geocentric distance (AU)
g: Solar Lyman-α g-factor (photons s⁻¹) at 1 AU
Q: Water production rates for each image (s⁻¹)
δQ: internal 1-sigma uncertainties

Figure 1 shows two versions of the variation of the water production rate as a function of time in days from perihelion. The diamonds give the individual single-image production rates; these are the water production rates using only data from each image. Because of the filling time of the hydrogen coma, the single-image production rate averages out the water production rate over the previous 1-3 days, depending on the observational geometry, and tends both to delay and to decrease the magnitude of any rapid changes in production rate like outburst timing and magnitude. For that reason we used the capability in the TRM to analyze the whole sequence of images together, the inversion of which extrapolates the water production rate from the vicinity of the nucleus by accounting for the dissociation times of $H_2O$ and OH as well as the transport time of H atoms from the inner to outer coma. These values are given in Table 2 and also shown in Figure 1 as the histogram values because they are averages over each day. These so-called deconvolved daily-average values show the outburst started a few days before being visible in the single-image values. However, it is worth noting the uncertainties are much larger because 2017 S3 was neither very bright nor had a very spatially extended coma. Because of the large uncertainties we continue to use the original single-image production rates in the rest of the quantitative physical analyses.

Adopting the method of Cowan and A'Hearn (1979) we have calculated the total active area from the production rate assuming a rapidly rotating sphere or spheres in the case of a distribution of sources, water sublimation and a visual geometric albedo of 0.04. The active area as a function of time is shown in Figure 2. The active area on day -45 with respect to perihelion was 5.5 $km^2$. If that corresponds to production from a single spherical nucleus without an extended icy-grain source, it implies a minimum radius of 660 m, assuming the comet is 100% water with only a small enough amount of dark material to make the surface albedo low. Of course it is possible that the nucleus was already surrounded by a halo of sublimating particles by that date, in which case the nucleus is likely much smaller.

Table 2. Deconvolved Daily-Average Water Production Rate of Comet 2017 S3 (PanSTARRS)

| T (days) | r (AU) | Δ (AU) | Q (10²⁸ s⁻¹) | δQ (10²⁸ s⁻¹) |
|---|---|---|---|---|
| -50.96 | 1.335 | 1.684 | 1.25 | 0.65 |
| -49.96 | 1.316 | 1.660 | 1.33 | 0.59 |
| -48.96 | 1.296 | 1.636 | 1.43 | 0.85 |
| -47.96 | 1.277 | 1.611 | 1.53 | 0.73 |
| -46.96 | 1.257 | 1.586 | 1.97 | 0.62 |
| -45.96 | 1.237 | 1.562 | 2.03 | 0.44 |
| -44.96 | 1.217 | 1.537 | 2.08 | 1.16 |
| -43.96 | 1.197 | 1.512 | 2.21 | 0.60 |
| -42.96 | 1.176 | 1.487 | 2.22 | 1.09 |
| -41.96 | 1.156 | 1.461 | 2.52 | 0.63 |
| -40.96 | 1.135 | 1.436 | 2.02 | 0.90 |
| -39.96 | 1.115 | 1.410 | 2.08 | 0.52 |
| -38.96 | 1.094 | 1.385 | 1.67 | 0.65 |
| -37.96 | 1.073 | 1.359 | 1.64 | 0.36 |
| -36.96 | 1.052 | 1.334 | 2.33 | 1.62 |
| -35.96 | 1.030 | 1.308 | 2.44 | 1.41 |
| -34.96 | 1.009 | 1.282 | 2.96 | 1.13 |
| -33.96 | 0.987 | 1.256 | 2.20 | 1.20 |
| -32.96 | 0.965 | 1.230 | 4.83 | 3.22 |
| -31.96 | 0.943 | 1.204 | 6.00 | 2.84 |
| -30.96 | 0.921 | 1.179 | 6.72 | 4.74 |
| -29.96 | 0.899 | 1.153 | 7.25 | 3.60 |
| -28.96 | 0.876 | 1.127 | 7.24 | 3.38 |
| -27.96 | 0.853 | 1.101 | 7.90 | 2.02 |
| -26.96 | 0.830 | 1.076 | 8.50 | 0.53 |
| -25.96 | 0.807 | 1.051 | 7.34 | 0.72 |
| -24.96 | 0.784 | 1.026 | 3.35 | 3.91 |
| -23.96 | 0.760 | 1.001 | 3.11 | 0.78 |
| -22.96 | 0.736 | 0.977 | 1.85 | 3.14 |
| -21.96 | 0.712 | 0.953 | 2.01 | 2.45 |
| -20.96 | 0.688 | 0.930 | 1.71 | 1.69 |
| -19.96 | 0.663 | 0.907 | 1.02 | 0.73 |
| -18.96 | 0.638 | 0.886 | 0.75 | 0.19 |
| -17.96 | 0.613 | 0.865 | 1.57 | 0.08 |
| -16.96 | 0.588 | 0.845 | 0.91 | 1.43 |
| -15.96 | 0.562 | 0.827 | 0.74 | 0.83 |
| -14.96 | 0.536 | 0.810 | 0.61 | 0.50 |

Notes to Table 2. ΔT: Time from perihelion on 2018 August 15.956 UT in days
r : Heliocentric distance (AU)
Δ: Geocentric distance (AU)
g: Solar Lyman-α g-factor (photons s⁻¹) at 1 AU
Q: Daily-Average Water production rates (s⁻¹) from the TRM
δQ: internal 1-sigma uncertainties

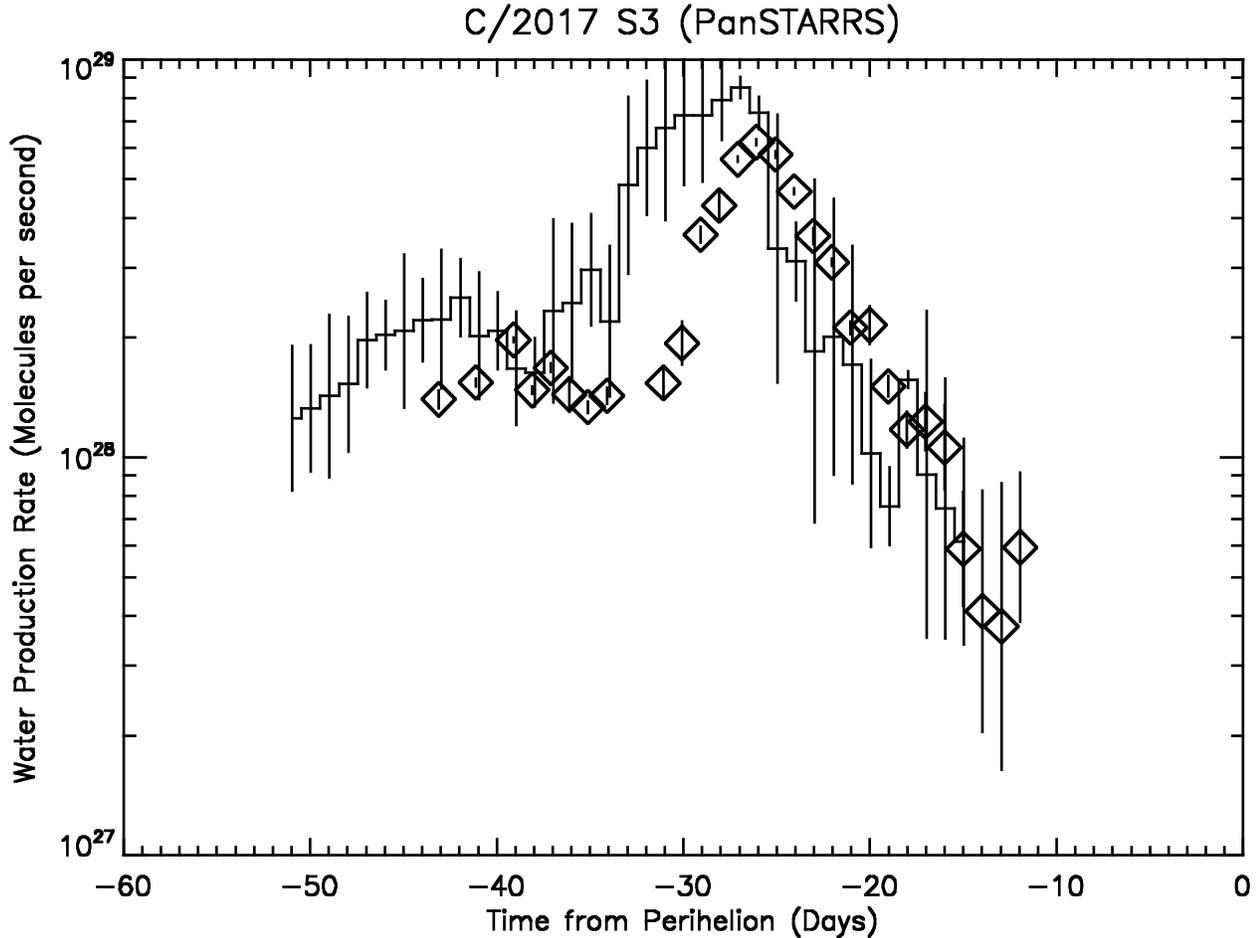

Figure 1. Single image (diamonds) and deconvolved daily (histogram) water production rates in comet C/2017 S3 (PanSTARRS) are plotted as a function of time from perihelion in days. The error bars correspond to the respective 1-sigma fitting uncertainties. The single image values, which represent the entire hydrogen content of the coma, delay the peaks of the outbursts as well as the disintegration decay by 1-3 days.

Integrating the total water mass loss over the observation period beginning on the day of the final outburst yields a value of 1.76 x 10$^9$ kg. If the total active area on day -45 does indicate sublimation from a single nucleus, then the total mass for a water-dominated nucleus of density 533 kg m$^{-3}$ (Pätzold et al. 2016), a radius of 660 m, which corresponds to the active area for a nearly pure water spherical nucleus, would be 6.4 x 10¹¹ kg. However, given the total water mass loss over the observation period, the radius of a single nearly pure water nucleus of the same bulk density would be only 92 m.

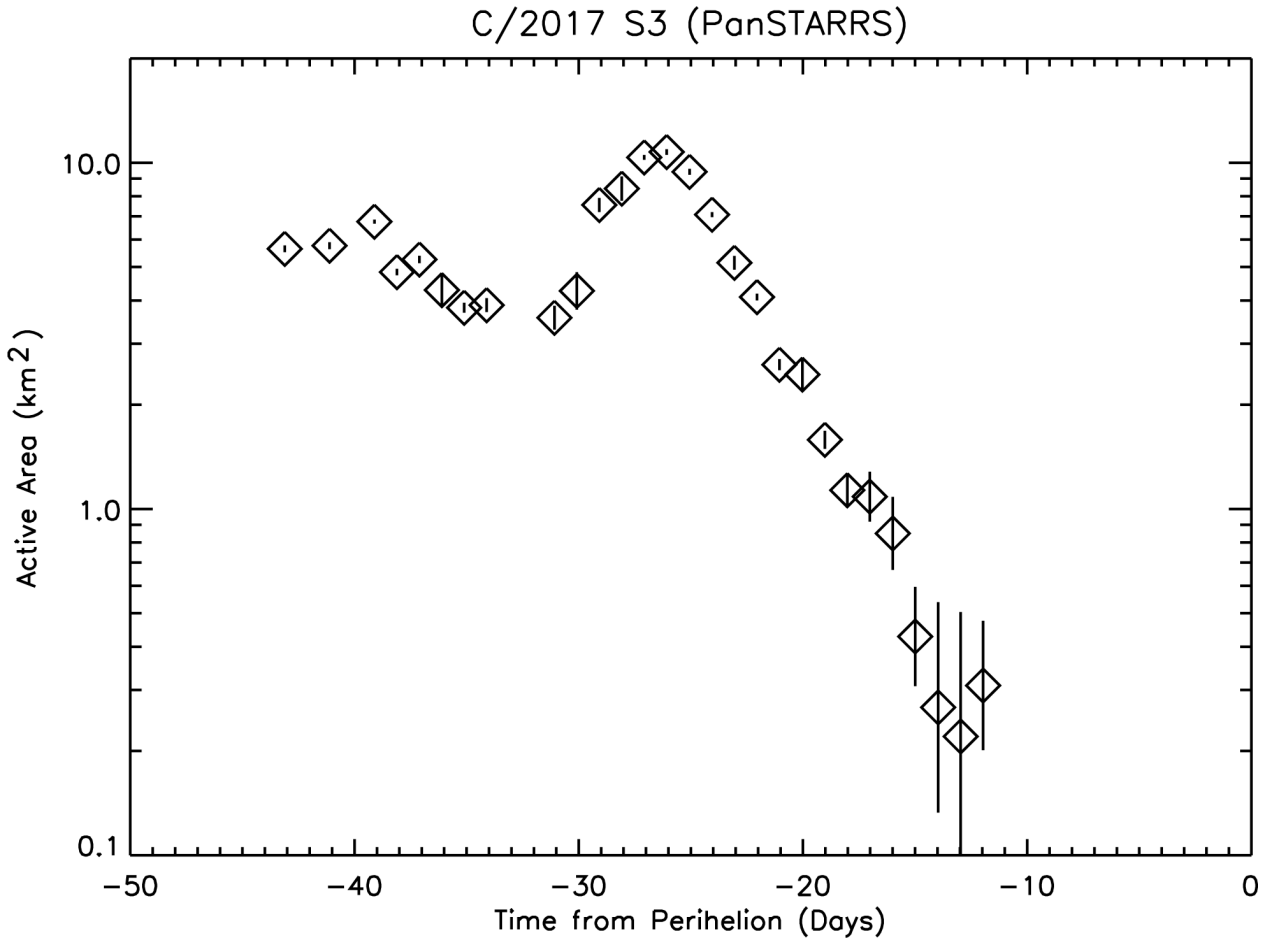

Figure 2. The active area (km²) of comet C/2017 S3 (PanSTARRS) as a function of time from perihelion in days. The diamonds give the active surface area calculated from the single-image water production rates. The rapid rotator method of Cowan and A'Hearn (1979) with a Bond albedo of 0.04 and perfect thermal emissivity was assumed.

Based on the level of production rate before the outbursts and the similar range of heliocentric distances, the size the active area of C/2017 S3 (PanSTARRS) is similar to that of C/2012 S1 (ISON) before its final runaway loss at much smaller heliocentric distances. However the total water mass loss of 2017 S3 through the end of its disintegration was almost an order of magnitude less than that of 2012 S1. In this sense, 2017 S3 appears to be much more similar to the other spectacularly lost comet 1999 S4 (LINEAR) which completely disintegrated at larger heliocentric distances of about 0.8 AU (Mäkinen et al. 2001). An alternative possibility is that 2017 S3 disintegrated into a number of larger chunks of the nucleus, which are only producing water by sublimation at a low rate.

1999 S4 was a smaller comet with a much lower production rate than 2017 S3, having a maximum production rate of $3.6 \times 10^{28}$ $s^{-1}$ at the peak of its last outburst compared with $8 \times 10^{28}$ $s^{-1}$. Furthermore, 1999 S4 lost $6.3 \times 10^8$ kg of water in its final disintegration or roughly 1/3 of the mass 2017 S3 lost. The active area of 1999 S4 was about 1 $km^2$ compared with 5.5 $km^2$. Based on the amount of dust scattered continuum after the break-up of 1999 S4, Farnham et al. (2000) suggested that it was originally a very ice-poor and dust-rich comet, so there was little ice to provide cohesiveness to keep it together or to continue sublimating for long after it disintegrated. They also suggested that 1999 S4 was already shedding icy particles long before the final disintegration. The case could very well be similar for 2017 S3 so that it was shedding sublimating icy particles from before the beginning of the SWAN observations meaning that the 660 m radius, or even a somewhat larger value after accounting for a more substantial dust/ice ratio, is much larger than the original size of the nucleus. Overall comet 2017 S3 might be simply a larger and more active version but otherwise very similar to 1999 S4. The various results regarding mass loss and active area size are summarized in Table 3.

It is clear that f=1 is only meant as a limiting case for only a very small amount of remaining refractory material containing little water. The value of 1/3 is at the average of the results obtained by the Rosetta gas instruments and dust instruments and seems reasonable for most comets, especially a dynamically new comet like C/2017 S3 (PanSTARRS). A value of 0.1 as implied by the Rosetta dust instruments is probably an extreme upper limit and would imply large values for the nucleus size and total mass.

Another comet that appeared to have an episode of icy-chunk/grain breakup as it approached perihelion was comet C/2014 Q1 (PanSTARRS) as described by Combi et al. (2018). It was not a dynamically new comet having an initial semi-major axis of 825 AU. With a perihelion distance of 0.314 AU, its average active area was 9.6 km$^2$ before perihelion assuming water sublimation. When it reached a heliocentric distance of about 0.7 AU, its production rate increased dramatically with an active area that increased by nearly an order of magnitude, indicating that it was continuously shedding material in the form of smaller particles, grains and chunks that subsequently sublimated. It continued with this elevated production through perihelion until it reached an outbound heliocentric distance again of about 0.7 AU, at which time the activity settled down more constant average active area of 4.9 km$^2$. Taken at face value this was consistent with 2014 Q1 lost roughly half its radius of material with a total water mass of $3.1 \times 10^{11}$ kg, much larger than either of these other disintegrated comets, but it did not completely disrupt or disintegrate.

Table 3. Estimates of a Spherical Nucleus Radius for C/2017 S3 (PanSTARRS)

| Water Mass Fraction of the Nucleus | f=1 | f=1/2 | f=1/3 | f=1/10 |
|---|---|---|---|---|
| Total Mass Lost from Day -41 to -15 (kg) | $1.76 \times 10^9$ | $3.52 \times 10^9$ | $5.28 \times 10^9$ | $1.76 \times 10^{10}$ |
| Radius of Spherical Nucleus from total Mass Loss (m) | 280 | 350 | 400 | 600 |
| Active area on Day = --50.96 (km$^2$) | 7.0 | 14.0 | 21.0 | 70.0 |
| Radius of Spherical Nucleus from Active Area on Day = -50.96 (m) | 700 | 1020 | 1250 | 4166 |

Following the approach Mäkinen et al. (2001) applied to the SWAN observations of 1999 S4 (LINEAR) we have calculated the distribution of sublimating icy particles produced during the final outburst on July 20 that would explain the rest of the water production rate variation. See the Appendix for the detailed quantitative description. Unlike the results for 1999 S4 that reproduced the production rate variation with a particle size distribution $N(R)\ dR \sim R^{-2.7}$, where R is the radius of the particles, we find for 2017 S3 (PanSTARRS) that we require a distribution much more heavily weighted toward the smaller particles with an exponent of $\sim$ -5.0. The size range is between 0.3 to 4 meters in radius assuming dark spherical particles. The disintegration activity is shown in Figure 3 along with the observed variation of water production rate following the final outburst.

While there could be a fundamental difference between particles released through normal comet activity of somewhat processed short-period comets like 67P and 103P and the more violent release during the total disruption of the nucleus of a dynamically new comet like 1994 S4 or 2017 S3, we can compare those here. The size distributions of particles in different comets vary widely, from the exponents of -2.7 and -3 found by Mäkinen et al. (2001) for 1999 S4 (Linear) and Bockelée-Morvan et al. (2001), respectively, to -3.5 found in 103P/Hartley 2 by Fougere et al. (2013) using the distribution of the extended offset of OH to values of -4.7 to -6.6 by Kelley et al. (2013) measuring spatial and flux distributions of particles directly. Fulle et al. (2016) reported the differential size distribution of dust particles in the Rosetta target comet 67P/Churyumov-Gerasimenko of -4 for sizes > 1mm by OSIRIS Agarwal et al. (2016) reported a differential size distribution power law exponent of -4.0 for particles > 9 cm. Ott et al. (2017) showed irregular dust mass distributions for large particles in a limited range of sizes. Measurements of small particles with sizes less than 1 mm with the Rosetta dust instrument GIADA report a changing power index of -2 just beyond 2 au to -3.7 at perihelion (Fulle et al. 2016).

Biver, Crovisier and Colom (2019, private communication) reported observing OH in comet C/2017 S3 (PanSTARRS) with the Nançay radio telescope on 20.5 July 2018 during the outburst and getting a production rate for OH of 1 - 1.6 x $10^{29}$ s$^{-1}$ not accounting for, and accounting for, quenching, respectively. This is somewhat larger than either the single-image (6.2 x $10^{28}$) or daily-average deconvolved (7.3 x $10^{28}$) found with SWAN, but it is possible with a smaller effective aperture and observing OH which has a shorter lifetime, the difference is due to the time averaging inherent in the SWAN measurement. They also reported that it was either not, or barely detectable before this on 14-15 July and again after the outburst on 22 July when the production rate was at least 3 times lower. They also reported visible range spectra that showed emissions of $C_2$ and CN with the $C_2$/CN ratio being higher than typical. They also reported an earlier outburst around July 3 to July 5, which coincides with the earlier SWAN observed outburst.

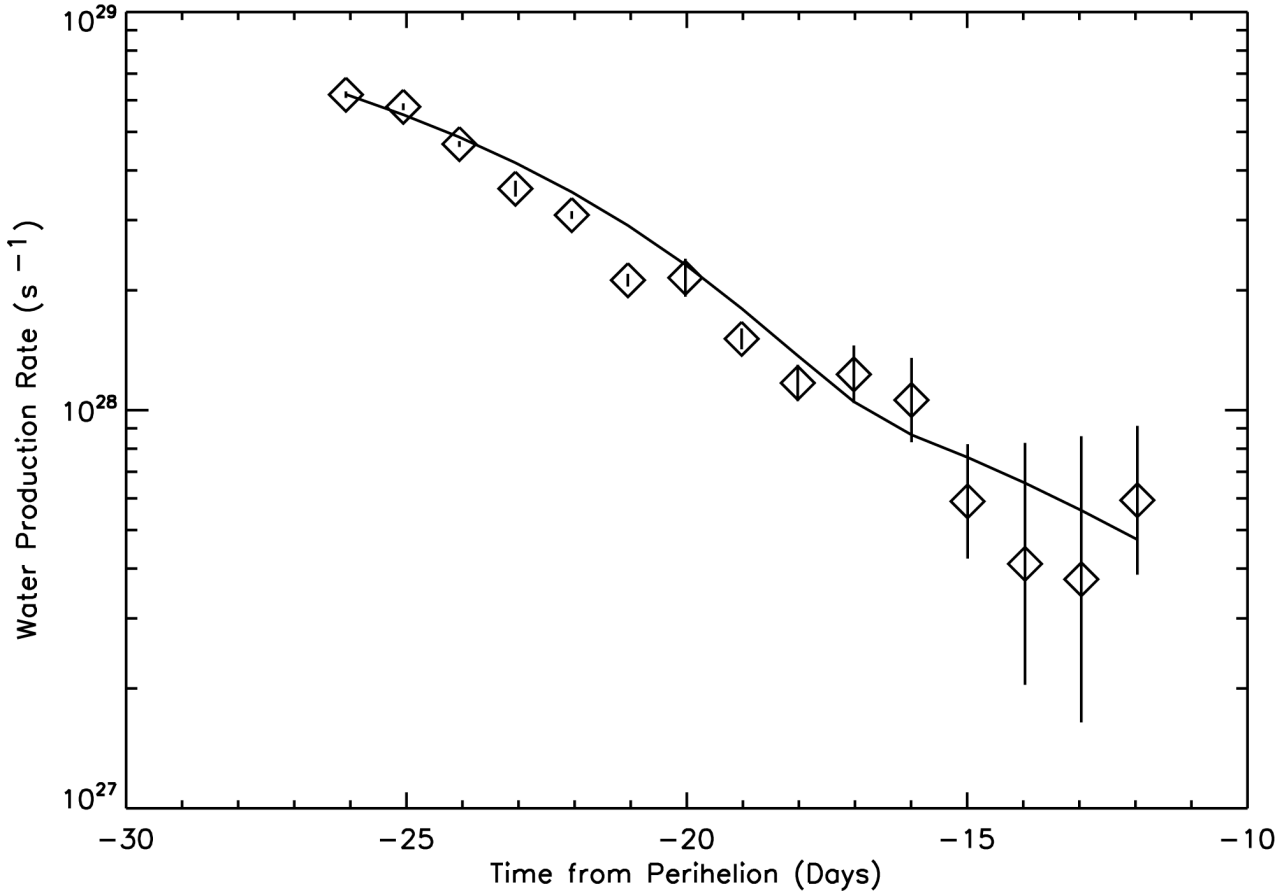

Figure 3. Water production rate of C/2017 S3 (PanSTARRS) compared with sublimation of particles after a fragmentation event on 20 July 2017. The points give the SWAN water production rates and the line gives the fragment sublimation model with a size distribution favoring the 0.3 m particles with a power law exponent of α=-5 and a total mass production of $1.76 \times 10^9$ kg. The particles range in size from 0.3 to 4 meters) with a bulk density of 533 kg m$^{-3}$ (Pätzold et al. 2019).

Biver (2019, private communication, http://www.lesia.obspm.fr/comets/lib/all-obs-table.php?Code=CK17S030&y1=1908&m1=01) reported that the comet was moving away from its predicted position suggesting strong non-gravitational acceleration. Images posted by their group show a typical concentrated nucleus condensation up until the outburst on July 19-20; but the last image taken on 3 August 2018 shows only a diffuse and somewhat elongated distribution with no detectable nucleus-centered condensation. The image is consistent with the SWAN results indicating complete disintegration of the nucleus and only continued spread and orbiting of the remaining debris cloud.

**Summary**

Comet C/2017 S3 (PanSTARRS) was observed by the SOHO SWAN H Lyα all-sky camera from 4 July to 4 August 2018. On its way to a perihelion distance of only 0.208 AU on 15 August, it underwent two outbursts, one only a couple of days after the first observation and a second on 20 July. After the 20 July outburst the water production rate dropped precipitously by more than a factor of 20 over the following two weeks despite the comet's heliocentric distance decreasing from 0.8 to 0.45 AU. This indicated that the comet was completely disintegrating with a behavior similar to that of comet 1999 S4 (LINEAR) in 2000. The water production rates and final disintegrated mass of 2017 S3 were a factor of almost 3 larger than 1999 S4. While the spectacular outbursts and disintegration of 1999 S4 were widely observed by both ground-based and space-based observatories, like HST, fully documenting the final dispersal of fragments (Weaver et al. 2001), 2017 S3 was unfortunately not well observed. However, the behavior documented by the SOHO SWAN observation indicates that 2017 S3 would have been similarly spectacular.

**Appendix**

We have adopted the fragment sublimation description from the results of the similarly disintegrated comet 1999 S4 (LINEAR) from the work of Mäkinen et al. (2001) and summarize it here. It is essentially similar to the contemporaneous one applied to 1999 S4 (LINEAR) radio observations by Bockelée-Morvan et al. (2001) who found a similar fragmentation particle distribution. For the collection of particles remaining after the last outburst, we adopt a particle size distribution of the form $N_i \sim R_i^{-\alpha}$ where $R$ is the particle size, $\alpha$ is the exponent of the size distribution and $N_i$ is the number of particles of size class *i*. For each size class the rate of change of production rate due to sublimation is given by

$$\frac{dQ_i}{dt} = -(9u\sqrt{\pi N_i}/\rho)[N_A F_S (1 - A_V)/Lr^2]^{3/2}[Q_i(t)]^{1/2}$$

where $Q_i$ is the production rate in s$^{-1}$, $u$ is the atomic mass unit, 1.66 x 10$^{-27}$ kg, $N_A$ is the Avagadro number, 6.022 x 10$^{23}$ mol$^{-1}$, $F_S$ is the solar constant, 1365 W m$^{-2}$, $A$ is the visual geometric albedo, assumed to be 0.04, $\rho$ is the bulk density of the nucleus taken to be 533 kg m$^3$ (Pätzold et al. 2019), $L$ is the latent heat of water ice, 50 kJ mol$^{-1}$, $r$ is the heliocentric distance in AU (see Table 1). As these particles sublimate (decay) in time they produce the water observed in the last two weeks of SWAN measurements. The total integrated production rate of water from day -26.079 to -11.964 is 1.76 x 10$^9$ kg. The best power law exponent, -5.0, is rather steep indicating a distribution highly peaked at the small end of the distribution of particles. Because of the daily time sampling and large aperture of SWAN particles we are not sensitive to particles much smaller than 0.3 m, so the size distribution corresponds to particle in the size range from 0.3 m to 4 m and with the steep slope found is heavily weighted to particles closer to 0.3 m. Chunks of this size are similar to those seen in 103P/ Hartley 2 by EPOXI (Hermalyn et al. 2013) and in 67P/Churyumov-Gerasimenko (Argawal et al. 2016) by Rosetta OSIRIS. The sum of the initial production rates for the particle size classes is 6.20 x 10$^{28}$ s$^{-1}$. This would be consistent with the drop in production rate seen by SWAN over 2-3 days after the outburst.

**Acknowledgements:** SOHO is an international mission between ESA and NASA. M. Combi acknowledges support from NASA grant 80NSSC18K1005 from the Solar System Observations Program. T.T. Mäkinen was supported by the Finnish Meteorological Institute (FMI). J.-L. Bertaux and E. Quémerais acknowledge support from CNRS and CNES. We obtained orbital elements from the JPL Horizons web site (http://ssd.jpl.nasa.gov/horizons.cgi). The composite solar Lya data was taken from the LASP web site at the University of Colorado (http://lasp.colorado.edu/lisird/lya/). We acknowledge the personnel that have been keeping SOHO and SWAN operational for over 20 years, in particular Dr. Walter Schmidt at FMI. We also acknowledge the support of R. Coronel by the Undergraduate Research Opportunity Program of the University of Michigan.

## References

A'Hearn, Michael F., Millis, Robert C., Schleicher, David G., Osip, David J., Birch, Peter V. 1995. Icarus, 118, 223.

Agarwal, J., A'Hearn, M.F., Vincent, J.-B., Gütter, C. and 40 colleagues. 2016. Mon. Not. Roy. Ast. Soc. 462, S78.

Bertaux, J.-L., Kyrölä, E., Quémerais, E., Pellinen, E. and 21 colleagues. 1995. Solar Physics 162, 403.

Bertaux, J.-L., Costa,J., Quémerais, E., Lallement, R., 6 colleagues. 1998. Planet. Space Sci. 46, 555.

Bockelée-Morvan, D., Biver, N., Moreno, R., Colom, P., Crovisier, J., Gérard, É., Henry, F., Lis, D.C., Matthews, H., Weaver, H.A., Womack, M. Festou, M.C. 2001. Science 292, 1339.

Combi, M.R., Fougere, N., Mäkinen, J.T.T., Bertaux, J.-L., Quémerais, E., Ferron, S. 2014. ApJ Lett 788:L7.

Combi, M.R., Mäkinen, T., Bertaux, J.-L., Quémerais and 3 coauthors. 2018. Icarus 300, 30

Combi, M.R., Mäkinen, T.T., Bertaux, J.-L., Quémerais, E., Ferron, S. 2019. Icarus, 317, 610.

Combi, M.R., Mäkinen, J.T.T., Bertaux, J-L. , Quémerais, E. 2005. Icarus 144, 191.

Combi, M.R., Smyth, W.H. 1988a. ApJ 327, 1026.

Combi, M.R., Smyth, W.H. 1988b. ApJ 327, 1044.

Cowan, J.J., A'Hearn, M.F. 1979. Moon and Planets, 21, 155.

Desvoivres, E., Klinger, J., Levasseur-Regourd, A.C., Jones, G.H. 2000. Icarus 144, 172.

Farnham, T.L. , Schleicher, D.G., Woodney, L.M., Birch, P.V., Eberhardy, C.A., Levy, L. Science 292, 1348.

Festou, M.C. 1981. Astron. Astrophys. 95, 69.

Fougere, N., Combi, M.R., Rubin, M., Tenishev, V. 2013. Icarus, 225, 688.

Fulle, M., Marzari, F., Della Corte, V., Fornasier, S. and 73 colleagues. 2016. ApJ, 821:19.

Harris, W.M., Combi, M.R., Honeycutt, R.K., Mueller, B.E.A., Scherb, F. 1997. Science 177, 676.

Hermalyn, B., Farnham, T.L., Collins, S.M., Kelley, M.S. and 8 colleagues. 2013. Icarus 222, 625.

Keller, H.U., Meier, R.R. 1976. Astron. Astrophys. 52, 273.

Lehky, M. Bacci, P., Maestripieri, M., Tesi, L .and 61 colleagues, 2017. Minor Planet Electronic Circ., No. 2018-Q62 (2018).

Lemaire, P., Emerich, C., Curdt,W., Schuehle, U., Wilhelm, K., 1998. Astron. Astrophys. 334, 1095.

Mäkinen, J.T., Bertaux, J-L., Combi, M.R., Quémerais, E. 2001, Science, 392, 1326.

Mäkinen, J.T., Combi, M.R. 2005. Icarus, 177, 217.

Ott, T., Drolshagen, E., Koschny, D., Güttler, C., and 49 colleagues. 2017. Mon. Not. Roy. Ast. Soc., 469, S276.

Pätzold, M., Andert, T.P., Hahn, M., Asmar, S.W.,  and 12 coauthors. 2016. Nature 530, 63.

Pätzold, M., Andert, T.P., Hahn, M. Barriot, J.-P., and 6 coauthors. 2019. Mon. Not. Roy. Ast. Soc. 483, 2337

Rodionov, A.V., Jorda, L., Jones, G.H., Crifo, J.F., Colas, F., Lecacheux, J. 1998, Icarus, 136, 232.

Schleicher, D.G., Millis, R.L., Osip, D.J., Lederer, S.M. 1998. Icarus 131, 233.

Wainscoat, R.J., Weryk R., Micheli, M., Woodworth, D. 2017. Central Bureau Electronic Telegrams, 4432, 1 (2017). Edited by Green, D. W. E.

Weaver, H.A., Sekanina, Z., Toth, I., Delahodde, C.E. and 17 colleagues. 2001 Science 392, 1329.

Yoshida, A. 2018. http:// http://www.aerith.net/comet/catalog/2017S3/2017S3.html.